\documentstyle[12pt,aaspp4]{article}

\begin{document}

\title{The Canada-UK Deep Submillimetre Survey: First Submillimetre Images, the
    Source Counts, and Resolution of the Background}

\author{Stephen Eales\altaffilmark{1}, Simon Lilly\altaffilmark{2}, 
Walter Gear\altaffilmark{3}, Loretta Dunne\altaffilmark{1}, J. Richard
Bond\altaffilmark{4}, Francois Hammer\altaffilmark{5}, Olivier
Le F\`evre\altaffilmark{6} and David Crampton\altaffilmark{7}}


\altaffiltext{1}{Department of Physics and Astronomy, Cardiff University,
P.O. Box 913, Cardiff CF2 3YB, UK}

\altaffiltext{2}{Department of Astronomy, University of Toronto, 60
St. George Street, Toronto, Ontario M5S 1A1, Canada}

\altaffiltext{3}{Mullard Space Science Laboratory, University
of London, Holmbury St. Mary, Dorking,
Surrey RH5 6NT, UK}

\altaffiltext{4}{CITA, University of Toronto, 60 St. George Street,
Toronto, Ontario M5S 1A1, Canada}

\altaffiltext{5}{DAEC, Observatoire de Paris-Meudon, 92195 Meudon, France}

\altaffiltext{6}{Laboratoire d'Astronomie Spatiale, Traverse de Siphon,
B.P.8  13376, Marseilles Cedex 12, France}

\altaffiltext{7}{Dominion Astrophysical Observatory, 5071 W. Saanich Road,
Victoria, BC, V8X 4M6, Canada}


\begin{abstract}

We present the first results of a deep unbiased
submillimetre survey carried
out at 450 and 850$\mu$m. We detected 12 sources at 850 $\mu$m, giving
a surface density of sources with $\rm S_{850 \mu m} > 2.8\ mJy$
of $\rm 0.49\pm0.16\ arcmin^{-2}$. The sources constitute
20-30\% of the
background radiation at 850$\mu$m and
thus a significant fraction of the entire
background radiation produced by stars. This implies, through the connection
between metallicity and background radiation, that a significant
fraction of
all the stars that have ever been formed were formed in objects
like those detected here.
The combination of their large contribution to the background
radiation and their extreme bolometric luminosities make these
objects excellent candidates for being proto-ellipticals. 
Optical astronomers have recently shown that the $UV$-luminosity
density of the universe increases by a factor of $\simeq$10 between $\rm
z=0$ and $\rm z=1-2$ and then decreases again at higher redshifts.
Using the results of a parallel submillimetre survey of the
local universe, we show that both the submillimetre
source density and 
background can be explained if the submillimetre
luminosity density evolves in a similar way to
the $UV$-luminosity density.
Thus, if these sources are ellipticals in the process of
formation, they may be forming at relatively modest redshifts.
\end{abstract}




%

\section{Introduction}

The recent confirmation of a submillimetre background (\cite{pug,fix,haus})
with approximately the same integrated intensity as the optical
background (\cite{poz,vog}) shows that to understand galaxy
formation and evolution it is crucial to take into account that part
of a galaxy's starlight that is absorbed by dust and re-radiated at
longer wavelengths.
The integrated background is the repository of all the 
stellar energy that has ever been emitted (with a
$1+z$ weighting factor), and so the contribution
of a particular class of galaxies to the background is a measure of
its importance in the star-formation history of the universe.
Determining the relative contribution to the background of different
galaxy classes
may be of particular importance for answering the question
of how ellipticals and spiral bulges formed.
The quantity of metals associated with nearby early-type galaxies (\cite{mike}) 
implies that early-type galaxies should be responsible for about
half the background (\S 4).
Nevertheless, optical surveys have failed to
find convincing evidence of proto-spheroids.
Possible explanations of this failure are (a) that when the stars formed
they were spread over a large volume of space before gradually coalescing
to form galaxies (explaining the absence of discrete sources), (b) that
the starbursts are hidden by dust. In both cases, these objects will
still be contributing to the background; in the latter case they
should be dominating the submillimetre background (\cite{bch,ee97}).

The prospects for resolving the submillimetre
background are good because of the commissioning of
the SCUBA submillimetre array (\cite{wayne,walt}) on the James Clerk
Maxwell Telescope.
Last year a group carried out a submillimetre
survey of two clusters, using the gravitational magnification by
the clusters as an aid for finding distant submillimetre sources.
They found four sources and concluded that optical surveys had
significantly underestimated the number of star-forming 
galaxies in the distant universe (\cite{sib}, SIB, \cite{blain}). We and 
others (\cite{barg,dave}) have now started 
`blind-field' surveys with SCUBA, which do not have the aid of gravitational
amplification but do not suffer from the uncertainties of the
cluster lens method. We are carrying out our survey in fields
used for the Canada-France Redshift Survey (\cite{lill95}) and our aim
is eventually to survey 200 arcmin$^2$ of the sky. In this paper
we present the first results of a survey of an area of 22 arcmin$^2$ and
analyse these using
the results of a parallel survey we are carrying out of the submillimetre
properties
of the local universe (\cite{loretta}). 
A second paper (\cite{pap2}, hereafter Paper II)
will discuss the optical counterparts to the distant submillimetre sources.

\section{Observations and Data Reduction}

We observed three fields simultaneously at 450 and 850 $\mu$m on 12 nights
during January and March 1998, using the same observing procedure
as SIB, with the exception that we chopped and nodded
30 arcsec EW.
We divided our
observations into integrations lasting about one hour, checking
the telescope pointing before and after each integration.
The optical depth during these nights
varied between 0.57 
and 1.65 at 450 $\mu$m and between
0.13 and 0.29 at 850 $\mu$m.
For the first field we simply made repeated
observations at the same position ($03^h\ 02^m\ 38.6^s\ 00^{\circ}\ 07'\ 40''$
[2000.0]), giving a total integration
time of 36.096 ks. For the second field we made eight integrations 
(21.504 ks in total) at
one position ($10^h\ 00^m\ 40.4^s\ 25^{\circ}\ 14'\ 20''$ [2000.0]) and 
four integrations (12.8 ks) at a second position
($10^h\ 00^m\ 38.4^s\ 25^{\circ}\ 14'\ 20''$ [2000.0])) 
27 arcsec away from the first. For the final field we made
observations in a mosaic pattern such that each point in the final map
was constructed 
from data received at nine locations within the instrument
field of view. Since the instrument is fixed in Naysmith coordinates, and
thus the bolometers move on the sky, this 
ensures that a point on the final map contains data from a large fraction
of the available 
bolometers, reducing the possibility of problems with individual
bolometers generating spurious sources.
We reduced our data in the standard way using the
SURF package (\cite{tim,sib}).
The final maps
are shown in Figs 1 \& 2.

\section{Source Extraction}

Since any real object is likely to be unresolved and will appear on 
a map as a positive source with two negative sources 30 arcsec on either
side, we used template analysis (\cite{pd}) to look for sources of this
kind on the 850-$\mu$m maps, using as a template a map of Uranus made
with the same 30-arcsec chop.
The negative emission is actually more complex than this because
the chop direction is only precisely EW at the beginning
of each integration, then remaining fixed at a constant
azimuthal angle (this causes the slight rotation
of the negative lobes visible in Fig. 2a), 
but this is a reasonable approximation.
After trimming
off the edge of each image, which is populated by noise artefacts due to
incomplete sampling of the sky near the array's edge,
we made both a map of the correlation
coefficient (CC), a measure of how closely the emission
resembles the template, and a map of the convolution of the image
with the template (Fig. 2). 
The noise on the template-convolved images is 0.56, 0.76, and 0.92
mJy per beam for the 3$^h$, 10$^h$, and 14$^h$ fields.
We included a source in our catalogue only if
both $\rm S/N > 3$ on 
the convolved map and $\rm CC > 0.4$, the latter criterion
reducing the chance of noise spikes, which do not have
the distinctive negative lobes, being included.

We checked the reality of our sources in two ways. First, we created
the negative of each map and confirmed that no sources were found
using the same selection criteria. Second, we generated
Monte-Carlo simulations of the fields, replacing 
the data for each bolometer (which is sampled every second) with
the output from a gaussian random-number generator with
the same standard deviation as for that bolometer (measured separately
for 
each one-hour integration). The simulations showed
that spurious sources tended to be concentrated towards the edges
of our fields as expected, 
and we used the simulations to define `reliable regions'
for each map (Figs 1 \& 2). The sources listed in Table 1 are
from these regions, and the simulations suggest that, on average, one
of these could be spurious. Because of the possibility that the rotation
of the chop direction
could affect fluxes estimated by template-fitting,
we also estimated fluxes by aperture photometry, finding
no significant difference for the two methods.

Fig. 3 shows the integral source counts from our survey. An
incompleteness analysis similar to that in SIB implies
we are missing 0, 19, and 48\% of the sources
with fluxes of 6, 4, and 3 mJy respectively.
Our source counts are consistent with those of the other small blank-field
surveys (\cite{barg,dave}) and of SIB.

\section{Discussion}

The sources we have detected at 850 $\mu$m account for 
19\% of the background emission at 850$\mu$m (Fig. 4), rising to
30\% if a correction is made for incompleteness, and 
$>$10\% of the 
background emission at 450$\mu$m.
Our four measurements and seven limits on the 450$\mu$m/850$\mu$m ratio
are consistent with the average ratio being similar to that
of the background emission, and thus consistent with the 850$\mu$m sources
being representative of the background at shorter submillimetre
wavelengths as well.
On the assumption that the sub-mm objects detected in our survey
make a negligible
contribution to the optical background (Paper II), these
objects account for $\sim$11\% of the total optical-near-IR plus submillimetre
background (\cite{dwek}), rising to $\sim$17\% if the incompleteness
correction is made. 

It is possible that this level of background emission
might be explained by AGN. If the black-hole masses measured in nearby galaxies
by HST
(\cite{ford}) are typical of galaxies in general, and if one makes the assumption
that 10\% of the mass accreting
into a black hole is turned into energy, then the AGN background
would be $\simeq$10\% of the background from stars. However, given the
uncertainties in this calculation, we will henceforth
make the more conservative assumption that we are observing emission from
dust heated by stars.

The
integrated background emission produced by a population at
a redshift $z$ is related to the smoothed out cosmic density
of processed material $<\rho (Z + \Delta Y) >$ produced by
these objects by
$ \int^{\infty}_0 I_{\nu} d\nu = {0.007 <\rho (Z + \Delta Y) > c^3 \over
4 \pi (1+z)} $ (\cite{pag}). Since our sources make up a significant
fraction of the background, they must represent
a population which has been responsible for a significant fraction
of all the star formation that has ever occured. 
In Paper II we show that these sources are similar
to the galaxies with huge far-infrared luminosities discovered with
IRAS (\cite{sand}). This combination of high individual bolometric
luminosities and a significant contribution to the background emission
is exactly what one expects for one mode of proto-spheroid
formation (\S 1), and it is hard to resist the conclusion that
these are these long-sought objects (see also \cite{barg,dave}). Note
that, with their
huge bolometric luminosities and implied star-formation rates,
it has long been suspected that ultra-luminous IRAS galaxies
might be ellipticals in the process of formation (\cite{wright}). What
has been lacking is direct evidence that objects like this at high
redshift are responsible
for a significant fraction of the star formation in the Universe.

An obvious question to ask is whether these objects are at $\rm z \sim 1-2$,
as are the galaxies that dominate the optical background (\cite{poz}),
or whether they are at higher redshifts, as one might suspect from
the ages of the stellar populations in nearby ellipticals (\cite{bow}).
We will address this question by making the null hypothesis that
the submillimetre luminosity density evolves in the same way as the
$UV$ luminosity density (\cite{sick}) and predicting the submillimetre
source counts and background. If the observed counts and background
exceed our predictions, this might be evidence for a population
of galaxies at $\rm z >> 1$. In Paper II we will examine this question
using the optical identifications.

The biggest problem in modelling the submillimetre source counts
and
background has been the lack
of information about the submillimetre properties of nearby galaxies.
Without this information, one is forced to start from
a local IRAS 60-$\mu m$ luminosity function and make an assumption
about the average far-infrared/submillimetre SED of galaxies, 
an extrapolation of a factor of over 10 in wavelength
(\cite{ee96}). 
To overcome this problem, we have been carrying out a parallel
submillimetre survey of the local universe (\cite{loretta}). By measuring
submillimetre fluxes for galaxies drawn from
statistically-complete samples of IRAS and optically-selected
galaxies, we can both determine the SED's for different types of
galaxy and (using accessible-volume techniques---\cite{avni}) 
the local submillimetre luminosity function.
As yet our local 850-$\mu$m luminosity function is only well
constrained at high submillimetre luminosities, which is acceptable
for making source-count predictions, since it is only the high-luminosity
sources which contribute at the flux densities of interest, but not
for estimating the background. Fig. 3 shows our source-count 
predictions. We implemented our null hypothesis in two ways:
(i) by having
the amplitude of the local luminosity function evolve with the same
form 
as the luminosity density; (ii) by having the luminosities of the sources
evolve with this form. Fig. 3 shows that with luminosity evolution, but
not with space-density evolution, the predicted counts match the
observed counts fairly well. 
If one assumes no evolution at all,
the observed source counts exceed the predicted counts by a large
factor, as noted by \cite{sib}.

For modelling the background we are thrown back on the IRAS 60-$\mu$m
luminosity function (\cite{law}), but we do have the advantage 
over previous studies of empirical results on the
far-IR-submillimetre SED's of galaxies. The 60, 100, and 850-$\mu$m
fluxes of IRAS galaxies are well fitted by a single-temperature
dust spectrum with a dust-emissivity index of 1 and
$T_d = 7.5 log_{10} L_{FIR} - 42.5$ over the range
$10.0 < log_{10} L_{FIR} < 12.5$, $L_{FIR}$ being defined as
in Lawrence et al. (\cite{loretta}). We do not claim that the dust-emissivity
index is 1 and that there is dust of only one temperature in these galaxies,
merely that this is a good empirical fit to the fluxes. Fig. 4 
shows predictions
using the IRAS luminosity function and this assumption about
galaxy SED's. The measured background is far in excess
of the no-evolution model, but the model in which the submillimetre
luminosity density evolves in the same way as the $UV$ luminosity
density matches the spectral shape and amplitude of the background 
remarkably well.

Our conclusion that the null hypothesis of evolution like
that in the optical waveband is acceptable (further supported
by the analysis in Paper II)
should be contrasted with 
recent claims to the contrary (\cite{blain,dave}).
The difference is
not in the data,
but in the assumptions made about the local
SED's. Blain et al., for example, assume $\rm T_d = 38\ K$ and $\rm n = 1.5$.
When compared with our assumptions about SED's, this makes a factor
of $\simeq$9 difference in intensity at 850 $\mu$m.

The level of the background can also be used to argue against
a major population
of dust-enshrouded proto-ellipticals
at $\rm z > 5$.
Let us assume
that all the stars in this hypothetical population
form instantaneously at
a redshift $z_F$ and that the SED from one of these
objects is the same as that of the archetypical local IRAS galaxy
Arp 220 (the precise SED is not important). This is enough
to define the {\it shape} of the background. The maximum amplitude
of this background is then set by the observed COBE background (Fig. 4).
We then subtract this maximal elliptical background from the
observed background to get a remnant background, which we assume
is from the spiral/irregular population. If we then assume
that late-type galaxies go through their period of maximum star
formation at $\rm z \sim 1$, we can then use the equation above and
the ratio of the two backgrounds to predict
the relative contributions to the present-day metal
abundance of the two populations. If $z_F = 5$ the dusty ellipticals
can produce $<$38\% of the metals of the other population
and if $z_F = 10$ this falls to $<$8\%. Observations suggest, however,
that the ratio of the global metal abundance associated with ellipticals
to that that associated with
spirals/irregulars lies between 1 and 6 (\cite{pag,mike}).
This discrepancy again suggests that ellipticals
form at relatively modest redshifts.

\acknowledgments

We are grateful to the many people that have contributed to the
development of SCUBA and to the JAC staff for their excellent support 
during this program, in particular Wayne Holland for
help in a multitude of areas. We also thank
Rob
Ivison for some useful advice.
Loretta Dunne is supported by a PPARC studentship.
Research by Dick Bond and Simon Lilly is supported by the Natural
Sciences and Engineering Research Council of Canada and by the
Canadian Institute of Advanced Research. The JCMT is operated 
by the Joint Astronomy
Center
on behalf of the UK Particle Physics and
Astronomy Research Council, the Netherlands Organization for Scientific
Research and the Canadian National Research Council.

\clearpage
 
\begin{deluxetable}{crrrrrr}
\footnotesize
\tablecaption{Source Catalogue. \label{tbl-1}}
\tablewidth{0pt}
\tablehead{
\colhead{Name} & \colhead{RA (2000.0)}   & 
\colhead{Dec (2000.0)}   & 
\colhead{S/N} &
\colhead{CC}  & 
\colhead{$\rm S_{850 \mu m}$ (mJy)} & 
\colhead{$\rm S_{450 \mu m}$ (mJy)} 
}
\startdata
CFRS03A & 03\ 02\ 36.27& 00\ 08\ 13& 6.1 & 0.61 & 3.4$\pm$0.6 & 17$\pm$7 \nl
CFRS10A & 10\ 00\ 38.19& 25\ 14\ 50& 4.2 & ... &  4.8$\pm$1.0 & 23$\pm$9 \nl
CFRS10B & 10\ 00\ 37.10 & 25\ 14\ 58& 5.0 & ... & 4.7$\pm$1.0 & 21$\pm$8 \nl
CFRS10C & 10\ 00\ 37.23 & 25\ 15\ 12& 3.8 & ... & 2.8$\pm$1.0 & $<$22 \nl
CFRS10D & 10\ 00\ 36.97 & 25\ 14\ 42& 3.6 & ... & 3.9$\pm$1.0 & $<$22 \nl
CFRS14A & 14\ 17\ 40.3& 52\ 29\ 08 & 8.8 & 0.78 & 8.8$\pm$1.1 & $<$31 \nl 
CFRS14B & 14\ 17\ 51.8& 52\ 30\ 29 & 6.3 & 0.64 & 5.2$\pm$0.9 & $<$31 \nl
CFRS14C & 14\ 17\ 33.6& 52\ 30\ 49 & 3.4 & 0.46 & 3.8$\pm$1.1 & $<$31 \nl
CFRS14D & 14\ 18\ 02.5& 52\ 30\ 53 & 3.6 & 0.47 & 3.2$\pm$0.9 & $<$31 \nl
CFRS14E & 14\ 18\ 02.6& 52\ 30\ 17 & 3.6 & 0.54 & 3.5$\pm$1.0 & $<$31 \nl
CFRS14F & 14\ 17\ 42.2& 52\ 30\ 27 & 3.1 & 0.43 & 2.7$\pm$0.9 & 20$\pm$11 \nl
CFRS14G & 14\ 17\ 29.7& 52\ 31\ 15 & 3.7 & 0.45 & 5.7$\pm$1.6 & ... \nl

\enddata

\tablecomments{Col. 1: name of source. Cols 2 \& 3: position (2000.0).
The positions were measured from the 850-$\mu$m maps,
except for those of CFRS03A, CFRS10A, and CFRS10B, which were measured
from the 450-$\mu$m maps. Col. 4: Signal-to-noise.
This is the peak signal on the map
produced by convolving the raw map with the template (see text) divided
by the noise on that map, except in the cases of the 10-hour sources,
where the signal-to-noise was measured from the raw map.
Col. 5: the correlation coefficient. The values for
the 10$^h$ sources are missing because template-fitting
was not used for this field. Col. 6: flux
at 850$\mu$m. These
were obtained by template-fitting,
except those for the 10-hour sources were
obtained by aperture photometry, using an aperture of diameter
20 arcsec. The errors include a
5\% calibration error. Col. 7: fluxes at 450$\mu$m. These
were obtained by aperture photometry,
using an aperture of diameter 12 arcsec, except that the flux for
CFRS03A was obtained by template-fitting. The errors include a calibration
error of 20\%. The limits are 3$\sigma$ upper limits. There is no measurement
for CFRS14G because the slightly smaller field-of-view of SCUBA
at 450 $\mu$m means that at this wavelength it falls in a noisy
edge region.}

\end{deluxetable}

\clearpage

\clearpage

\figcaption[]{(Plate 1) Images at 850 $\mu$m of the 
3$^h$ (1a) and the 10$^h$ (1b) fields. The beam of the telescope
has a FWHM of 14 arcsec at 850 $\mu$m.
The 3$^h$ image has been convolved with a 14-arcsec gaussian (FWHM),
the 10$^h$ image with an 8-arcsec gaussian. The contour lines are drawn
at $\pm$2, 3, 4, 5....$\sigma$. The dashed line shows the region which is
believed to be free of edge artefacts and to have reasonably uniform
noise properties. This region is
$\simeq$110 arcsec across for the 3$^h$ field and $\simeq$120 across
(in the vertical direction) for the 10$^h$ field. \label{Fig. 1 (Plate 1)}}

\figcaption[]{(Plate 2) Images at 850 $\mu$m of the 14$^h$ field. The upper
image (2a) has been convolved with a 14-arcsec gaussian, the lower image
(2b) has been convolved with the source template (see text). 
Both images have been divided by the square root of a variance array
produced from a Monte-Carlo simulation of the field, which 
corrects for uneven noise across the images.
The contour lines 
are drawn at $\pm$2, 3, 4, 5....$\sigma$.
As the convolution with the
template
uses information from both the positive and negative `lobes',
there are more prominent sources in 2b than in 2a,
although one has to beware spurious
sources being created 60 arcsec EW from a bright source (e.g. source
Z).
The dashed line in the lower
image shows the region which is 
believed to be free of edge artefacts.
This is $\simeq$390 arcsec across in the horizontal 
direction. \label{Fig. 2 (Plate 2)}} 

\figcaption[]{Integral source counts at 850$\mu$m from our survey
(dots) and other surveys: triangle (\cite{dave}),
circle (\cite{barg}) and cross (\cite{sib}). The errors on our
counts are
Poisson errors and are not independent. No correction has been 
to any of the counts for incompleteness, although the incompleteness
for the other two blind-field surveys is claimed to be low. 
The predictions were made as described in the test. The key is as follows:
no evolution, $\rm \Omega_0 = 1$---squares; density evolution,
$\rm \Omega_0 = 0$---dots in circles; density evolution,
$\rm \Omega_0 = 1$---circles; luminosity evolution,
$\rm \Omega_0 = 0$---stars; luminosity evolution,
$\rm \Omega_0 = 1$---triangles. \label{Fig. 3}} 

\figcaption[]{The dashed line and the points show, respectively,
the submillimetre background estimated from the results of the 
FIRAS (\cite{fix}) and DIRBE (\cite{haus}) experiments on COBE.
The two lower limits show the limits on the background produced
by simply adding up the number of sources that we observe.
The solid line and the dotted lines show backgrounds predicted
using an IRAS luminosity function and information on the 
SED's of galaxies from our nearby galaxy survey.
The solid line is for no evolution, the dotted line for
a model in which the submillimetre luminosity density evolves like
the UV luminosity density. The two other lines are the backgrounds
predicted for a model in which all the stars in ellipticals form
instantaneously at a redshift $z_F$, with the stars being hidden
by dust. The predicted backgrounds have been normalized so that they
are the maximum backgrounds consistent with the FIRAS measurement.
The higher curve is for $z_F = 5$, the lower for 
$z_F = 10$.\label{Fig. 4}}


\begin{thebibliography}{}

\bibitem[Avni and Bahcall 1980]{avni} Avni, Y. \& Bahcall, J.N. 
    1980, \apj, 235, 694.
\bibitem[Barger et al. 1998]{barg} Barger, A.J. et al., Nature, in press.
\bibitem[Blain et al. 1998]{blain} Blain, A.W., Smail, I., Ivison,
    R.J. \& Kneib, J-P. 1998, \mnras, submitted (astro-ph 9806063).
\bibitem[Bond et al. 1986]{bch} Bond, J.R., Carr, B.J. \& Hogan,
    C.J. 1986, \apj, 306, 428.
\bibitem[Bower et al. 1992]{bow} Bower, R.G., Lucey, J.R. \& Ellis,
    R.S. 1992, \mnras, 254, 601.
\bibitem[Dunne et al. 1998]{loretta} Dunne, L. et al., in preparation.
\bibitem[Dwek et al. 1998]{dwek} Dwek, E. et al. 1998, \apj, in press.
\bibitem[Eales and Edmunds 1996]{ee96} Eales, S.A. \& Edmunds, M.G. 1996.
    280, 1167.
\bibitem[Eales and Edmunds 1997]{ee97} Eales, S.A. \& Edmunds, M.G. 1997,
    \mnras, 286, 732.
\bibitem[Edmunds and Phillipps 1997]{mike} Edmunds, M.G. \& Phillipps,
    S. 1997, \mnras, 292, 733.
\bibitem[Fixsen et al. 1998]{fix} Fixsen, D.J. et al. 1998, \apj, in press.
\bibitem[Ford et al. 1998]{ford} Ford, H.C., Tsvetanov, Z.I., Ferrarese, L.
    \& Jaffe, W. 1998, The Central Regions of the Galaxy and Galaxies, IAU
    184, in press (astro-ph 9711299).
\bibitem[Gear et al. 1998]{walt} Gear, W. et al., in preparation.
\bibitem[Hauser et al. 1998]{haus} Hauser, M.G. et al. 1998, \apj, in press.
\bibitem[Holland et al. 1998]{wayne} Holland, W.S. et al. 1998, Advanced
    Technology MMW, Radio and Terahertz Telescopes, ed. Phillips, T.,
    Proc SPIE 3357, 1998.
\bibitem[Hughes et al. 1998]{dave} Hughes, D.H. et al. 1998, Nature, in press.
\bibitem[Jenness 1997]{tim} Jeness, T. 1997, SURF - SCUBA user reduction
    facility. {\it Starlink User Note 216.1}.
\bibitem[Lawrence et al. 1986]{law} Lawrence, A., Walker, D.,
    Rowan-Robinson, M., Leech, K.J. \& Penston, M.V. 1986, \mnras, 219, 687.
\bibitem[Lilly et al. 1998]{pap2} Lilly, S., Eales, S.A., Gear, W., 
    Hammer, F., Le F\`evre, O., Crampton, D., Bond, J.R. \& Dunne, L., 
    in preparation.
\bibitem[Lilly et al. 1995]{lill95} Lilly, S., Le F\`evre, O.,
    Crampton, D., Hammer, F. \& Tresse, L. 1995, \apj, 455, 50.
\bibitem[Madau et al. 1998]{sick} Madau, P., Della Valle, M. \&
    Panagia, N. 1998, \mnras, 279, L17.
\bibitem[Pagel 1997]{pag} Pagel, B. 1997, Nucleosynthesis and
    Chemical Evolution of Galaxies (CUP).
\bibitem[Philips and Davies 1991]{pd} Phillips, S. \& Davies, J. 1991,
    \mnras, 251, 105.
\bibitem[Pozzetti et al. 1998]{poz} Pozzetti, L., Madau, P., Zamorani, G.,
    Ferguson, H.C. \& Bruzual, G.A. 1998, \mnras, submitted (astro-ph 9803144).
\bibitem[Puget et al. 1996]{pug} Puget, J-L. et al. 1996, A\&A, 308, L5. 
\bibitem[Sanders and Mirabel 1996]{sand} Sanders, D.B. \& Mirabel,
    I.F. 1996, \araa, 74, 749.
\bibitem[Smail et al. 1997]{sib} Smail, I., Ivison, R.J. and
    Blain, A.W. 1997, \apjl, 490, L5.
\bibitem[Vogeley 1998]{vog} Vogeley, M.S. 1998, \apj, 
    submitted (astro-ph 9711209).
\bibitem[Wright et al. 1990]{wright} Wright, G.S., James, P.A., Joseph, R.D.
   \& McLean, I.S. 1990, Nature, 344, 417.


\end{thebibliography}
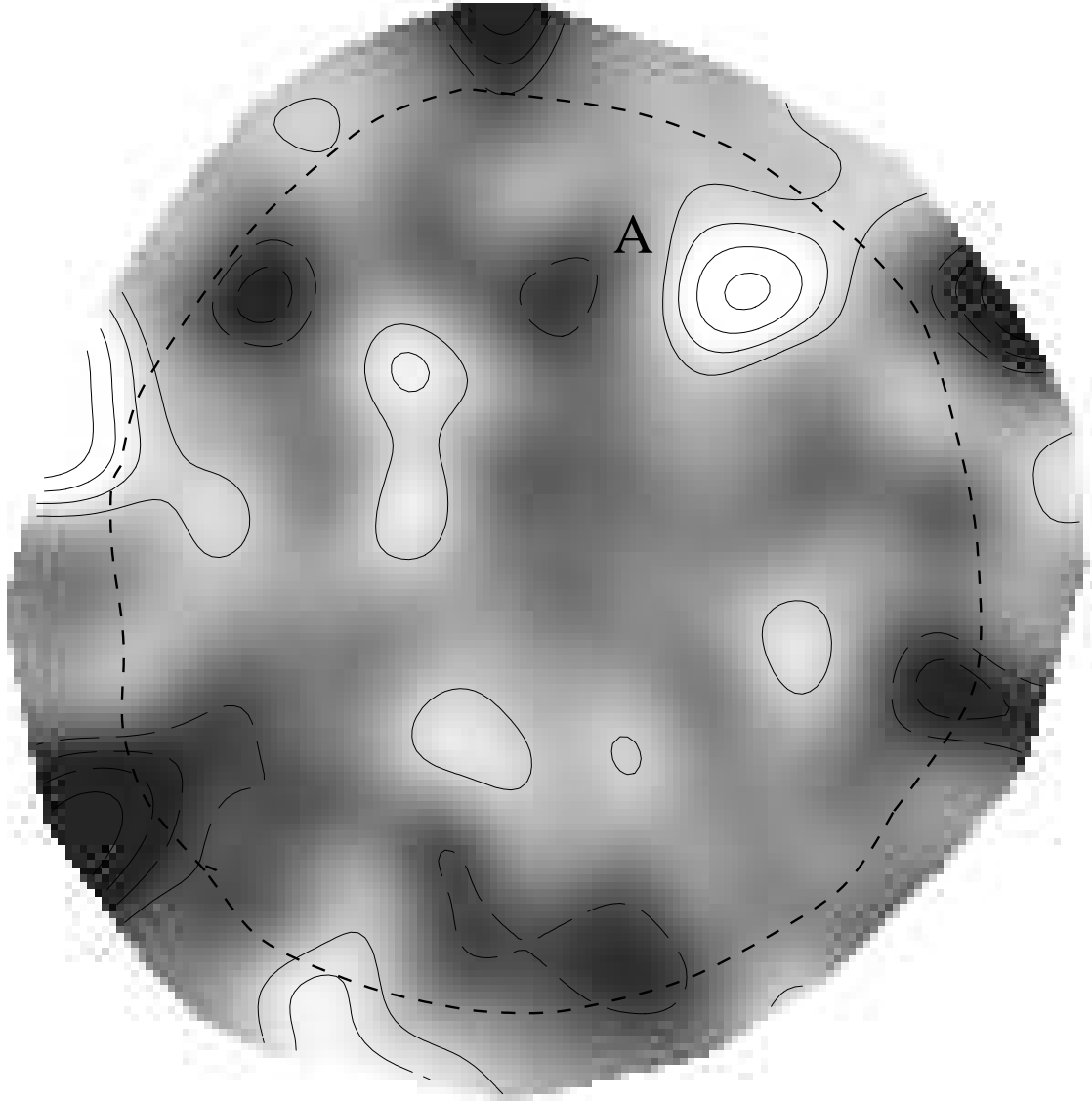
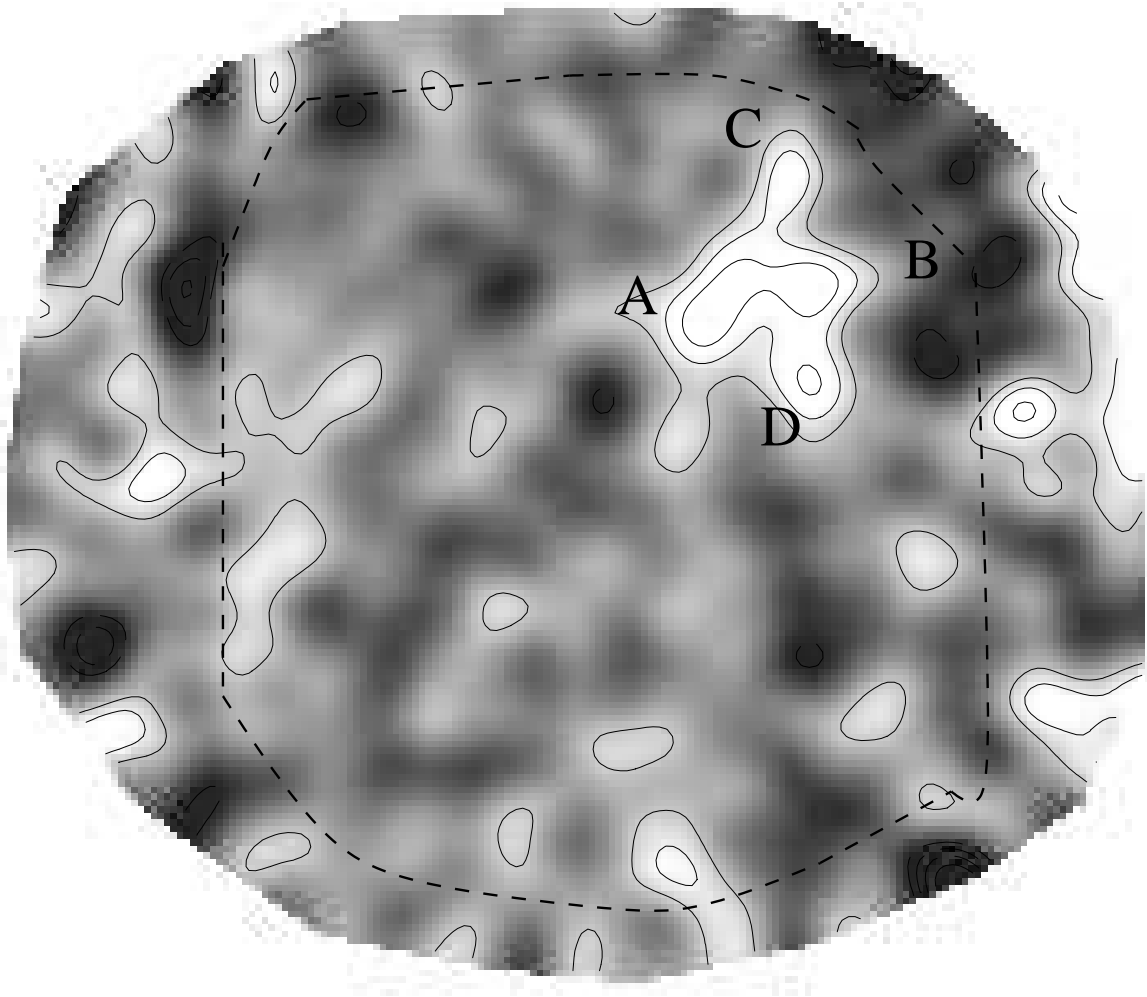
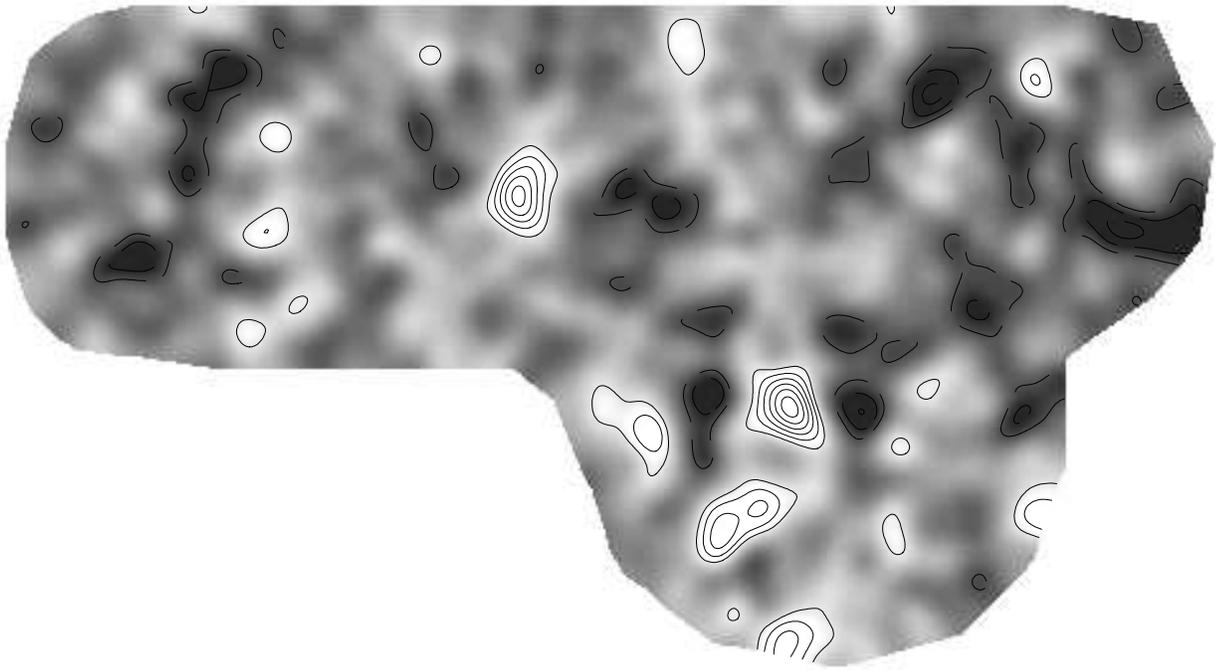
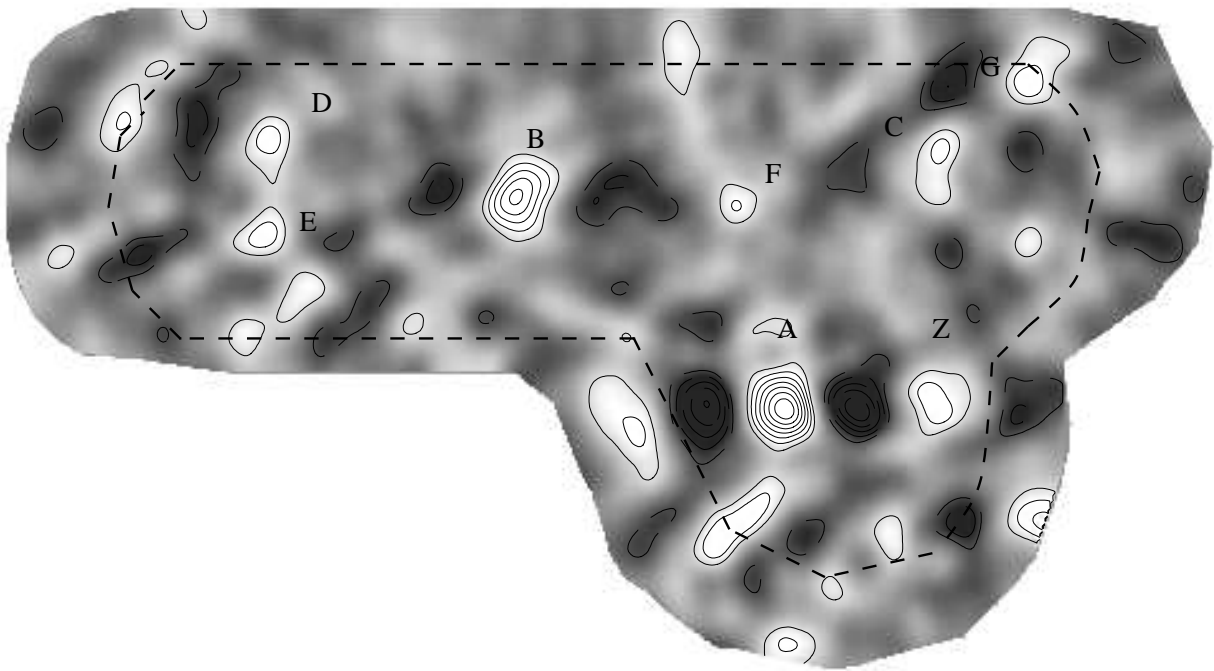
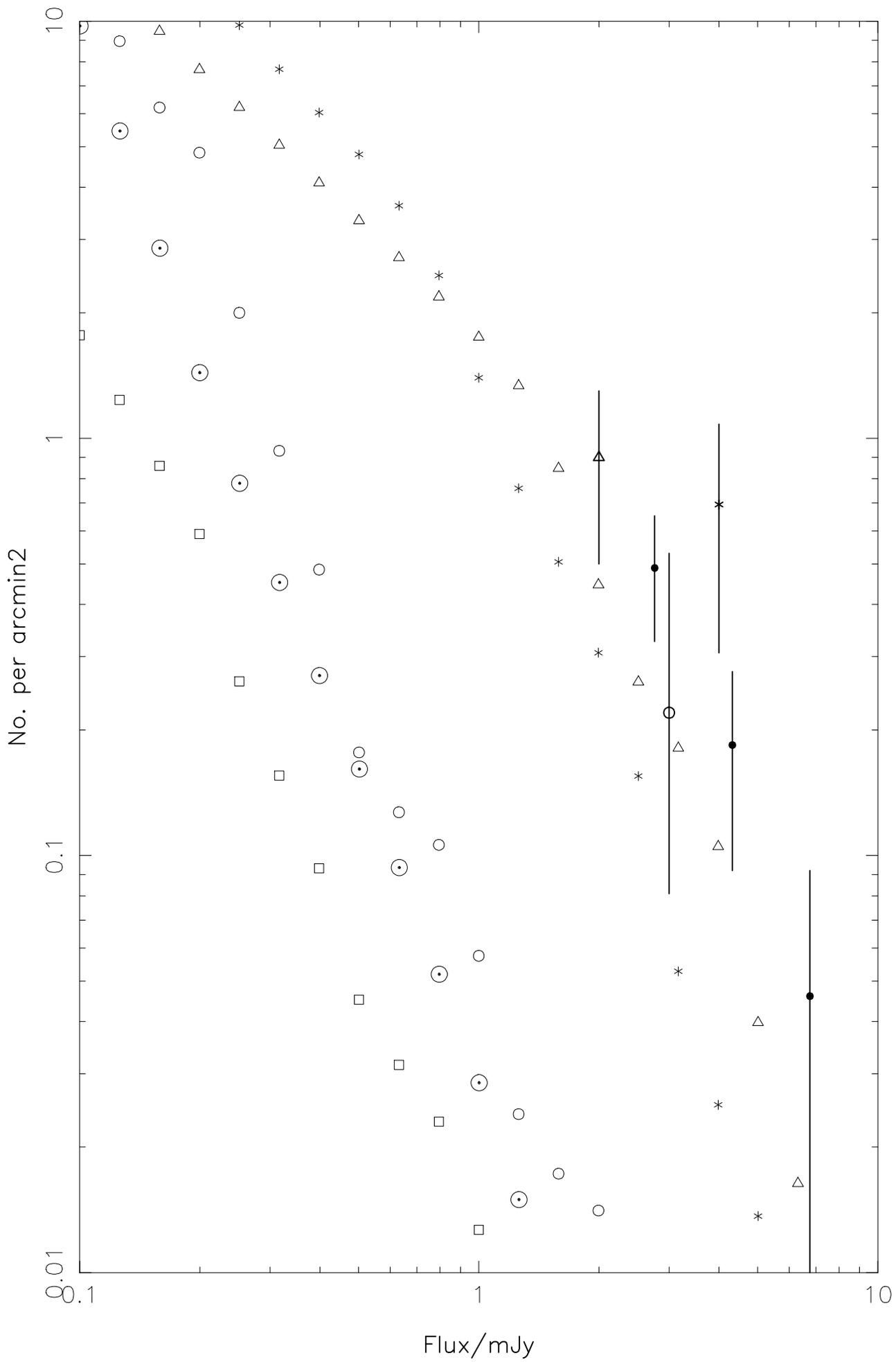
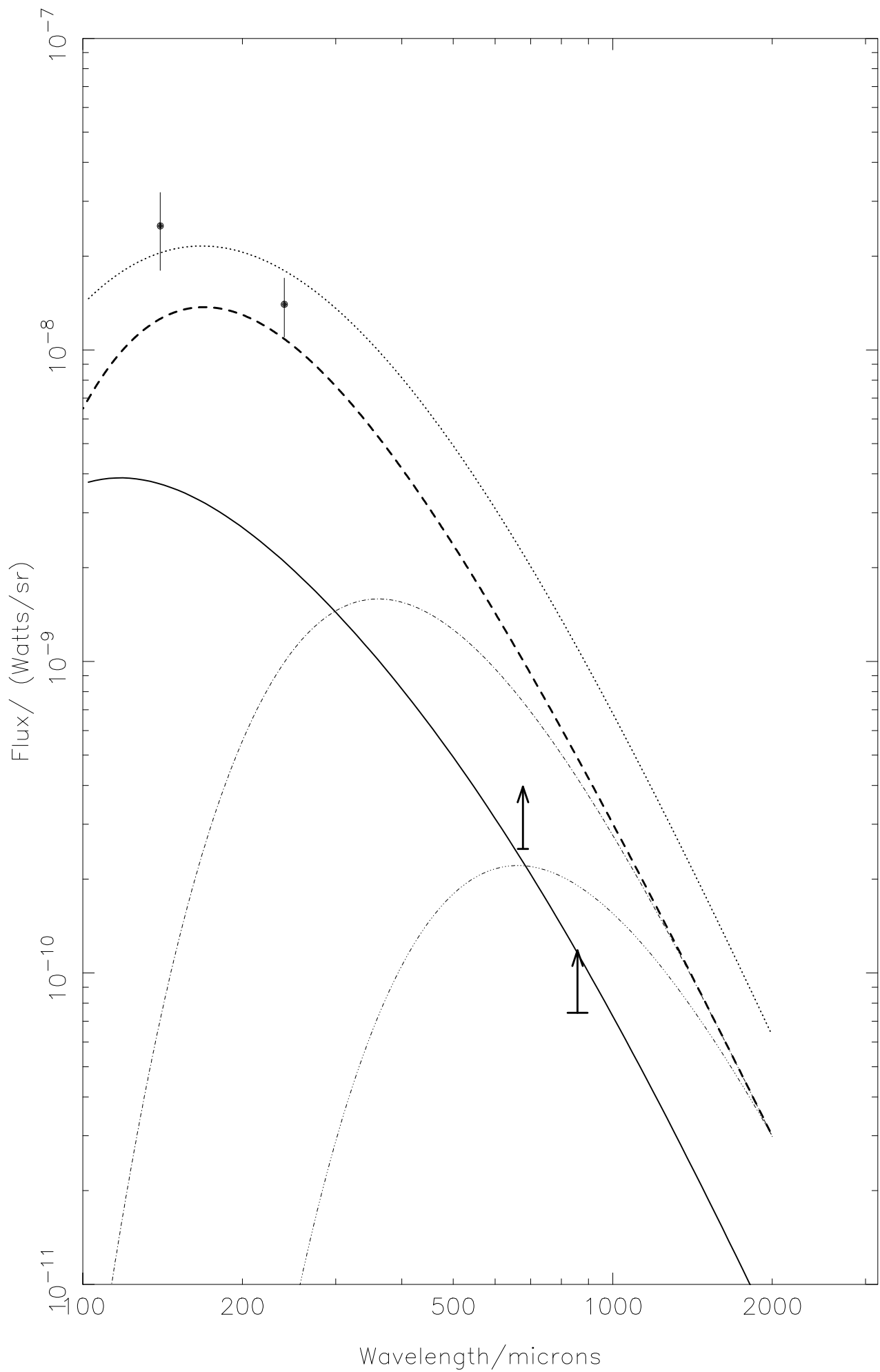
\end{document}